\documentclass{article}
\usepackage{lineno}
\usepackage[utf8]{inputenc}
\usepackage{graphicx}
\usepackage{authblk}
\usepackage{geometry}
\title{Spatial variation in the basic reproduction number of COVID-19: A systematic review}
\author[1,*,a]{RN Thiede}
\author[2,b]{N Abdelatif}
\author[1,c]{IN Fabris-Rotelli}
\author[3,d]{R Manjoo-Docrat}
\author[4,e]{J Holloway}
\author[2,f]{C Janse van Rensburg}
\author[3,4,g]{P Debba}
\author[4,h]{N Dudeni-Tlhone}
\author[5,i]{Z Kimmie}
\author[4,j]{A le Roux}

\affil[1]{Department of Statistics, University of Pretoria}
\affil[2]{Biostatistics, South African Medical Research Council}
\affil[3]{Department of Statistics and Actuarial Science, University of the Witwatersrand}
\affil[4]{Council for Scientific and Industrial Research, Pretoria, South Africa}
\affil[5]{Foundation for Human Rights, South Africa}

\affil[*]{Corresponding author: renate.thiede@gmail.com}
\affil[a]{ORCiD ID: 0000-0003-0934-5374}
\affil[b]{ORCiD ID: 0000-0002-3185-1284}
\affil[c]{ORCiD ID: 0000-0002-2192-4873}
\affil[d]{ORCiD ID: 0000-0003-2039-0440}
\affil[e]{ORCiD ID: 0000-0003-0948-4796}
\affil[f]{ORCiD ID: 0000-0002-6539-7833}
\affil[g]{ORCiD ID: 0000-0003-4870-988X}
\affil[h]{ORCiD ID: 0000-0002-8853-3121}
\affil[i]{ORCiD ID: 0000-0002-6392-0619}
\affil[j]{ORCiD ID: 0000-0002-9214-5076}

\date{November 2020}

\begin{document}

\maketitle

Funding: This work is supported by the NRF-SASA (National Research Foundation-South African Statistical Association) Crisis in Statistics Grant. 

\newpage

\begin{abstract}
OBJECTIVES: Estimates of the basic reproduction number ($R_0$) of COVID-19 vary across countries. This paper aims to characterise the spatial variability in $R_0$ across the first six months of the global COVID-19 outbreak, and to explore social factors that impact $R_0$ estimates at national and regional level.

METHODS: We searched PubMed, LitCOVID and the WHO COVID-19 database from January to June 2020. Peer-reviewed English-language papers were included that provided $R_0$ estimates. For each study, the value of the estimate, country under study and publication month were extracted. The median $R_0$ value was calculated per country, and the median and variance were calculated per region. For each country, the Human Development Index (HDI), Sustainable Mobility Index (SMI), median age, population density and development status were obtained from external sources.

RESULTS: A total of 81 studies were included in the analysis. These studies provided at least one estimate of $R_0$, along with sufficient methodology to explain how the value was calculated. Values of $R_0$ ranged between 0.48 and 14.8, and between 0.48 and 6.7 when excluding outliers.

CONCLUSIONS: This systematic review provides a comprehensive overview of the estimates of the basic reproduction number of COVID-19 globally and highlights the spatial heterogeneity in $R_0$. The value was generally higher in more developed countries, and countries with an older population or more sustainable mobility. Countries with higher population density had lower $R_0$ estimates. For most regions, variability in $R_0$ spiked initially before reducing and stabilising as more estimates became available.


\end{abstract}

\section{Introduction}

In 2020, the world observed the spread of the COVID-19 pandemic, which was of an unprecedented nature. The World Health Organisation released a statement on 9 January 2020\footnote{https://www.who.int/china/news/detail/09-01-2020-who-statement-regarding-cluster-of-pneumonia-cases-in-wuhan-china} reporting that a novel respiratory disease had been identified by Chinese authorities, originating in Wuhan. This disease was identified as 2019-nCov in early January, and spread rapidly outside of Wuhan and into the rest of the world. By 17 January, China recorded 62 cases, and 3 travelers had exported the disease, with two diagnosed in Thailand and one in Japan \cite{gralinski2020return}. At the time of writing, 220 countries have been affected by the disease, with the total death toll estimated at over 1 500 000\footnote{https://www.worldometers.info/coronavirus/ Accessed on 7 December 2020.}. This highlights the importance of understanding all aspects of the disease in order to combat it effectively.

The basic reproduction number ($R_0$) is of particular importance in epidemiological modelling \cite{ridenhour2018unraveling}. This is the average number of susceptible individuals that are infected by a single infected individual \cite{dietz1993estimation}. It is best measured in the early phase of the outbreak, before control measures have had time to take effect, and when most of the population is susceptible. Estimation of $R_0$ has proved complex \cite{petrosillo2020covid}.  $R_0$ estimation depends on the extent to which the disease characteristics are understood, which is challenging in the case of a novel disease. Assumptions must also be made about the ways in which people come into contact with one another. These assumptions and uncertainties limit the precision and accuracy of models used to determine $R_0$. Previous reviews have determined that values of the estimate range between 0.6 and 14.8 \cite{xie2020insight,salzberger2020epidemiologie}. The reviews of \cite{park2020systematic} and \cite{liu2020reproductive} determined $R_0$ to be between 1 and 7, while \cite{lv2020coronavirus,alimohamadi2020estimate,he2020estimation} found $R_0$ estimates to be between 2 and 4. The  review of \cite{rahman2020basic} states that estimates appear to be stabilising between 2 and 3. It should be noted that these $R_0$ estimates were all determined at the beginning of the pandemic, and do not reflect changes in the disease over time.

Reasons for the great variability in $R_0$ estimates remain an open question. The review of \cite{liu2020reproductive} indicated that mathematical models and stochastic models gave different values for $R_0$, however, \cite{alimohamadi2020estimate} stated that heterogeneity in the estimates was not influenced by the estimation methods chosen. The data used to estimate $R_0$ introduces another source of variability. This is supported by the review of \cite{he2020estimation}, which states that $R_0$ varies from place to place, but does not describe the variation. While comprehensive, none of the above reviews investigated how the estimates of $R_0$ differ across countries or regions. 

This systematic review aims to provide a comprehensive overview of the estimates of the basic reproduction number of COVID-19 globally and to highlight the spatial heterogeneity in the estimates over the first half of 2020. As $R_0$ is measured at the beginning of the pandemic, changing $R_0$ values do not reflect the progress of the disease, but instead illustrate a growing understanding of the disease characteristics over time. We further investigate possible factors that influence this spatial variability, namely their economic development status as classified by the World Bank, their human development index (HDI) as an indicator of social development, social mobility index (SMI) as a measure of mobility, median age and population density. Finally, an illustrative example considering a model specific to South Africa is given to guide the discussion on spatial heterogeneity in $R_0$.

\section{Materials and Methods}
\subsection*{Inclusion criteria}
Only English peer-reviewed research papers were included. These had to provide an estimate of $R_0$, and had to provide enough information in the methodology to determine how $R_0$ was calculated. Reviews, conference papers and other sources were not considered directly in our review, but were consulted as secondary sources.

\subsection*{Search strategy}
The search included papers published between 31 December 2019 and 30 June 2020, allowing for the initial stages of the disease spreading internationally. The WHO COVID-19 database, PubMed and LitCovid were searched for English-language papers including the terms “Coronavirus”, “COVID-19”, “SARS-CoV-2”, or “2019-nCoV” in combination with the phrases “reproduction number” or “reproductive number”.  Two reviewers (RT and NA) independently evaluated the eligibility of the studies obtained from the literature searches. 
All articles yielded by the database searches were screened by their titles and abstracts to obtain studies that met the inclusion criteria. After this, full-text screening was performed. In cases of uncertainties, agreement was reached through a third reviewer (IFR).

\subsection*{Data extraction}
Data extracted included the country for which $R_0$ was estimated, whether actual data or simulated data was used, the value of $R_0$ as well as its 95\% confidence interval and whether a spatial component was considered in the estimation of $R_0$. Data was extracted independently by two reviewers (RT and NA) and in the case of no consensus, IFR was consulted. Risk of bias was not considered in this study because no meta-analysis was done. 

\subsection*{Ethics statement}
This study was a systematic review which did not employ any human or animal subjects.

\section{Results}

\subsection*{Description of included articles}

\begin{figure}
    \centering
    \includegraphics[width = 10cm]{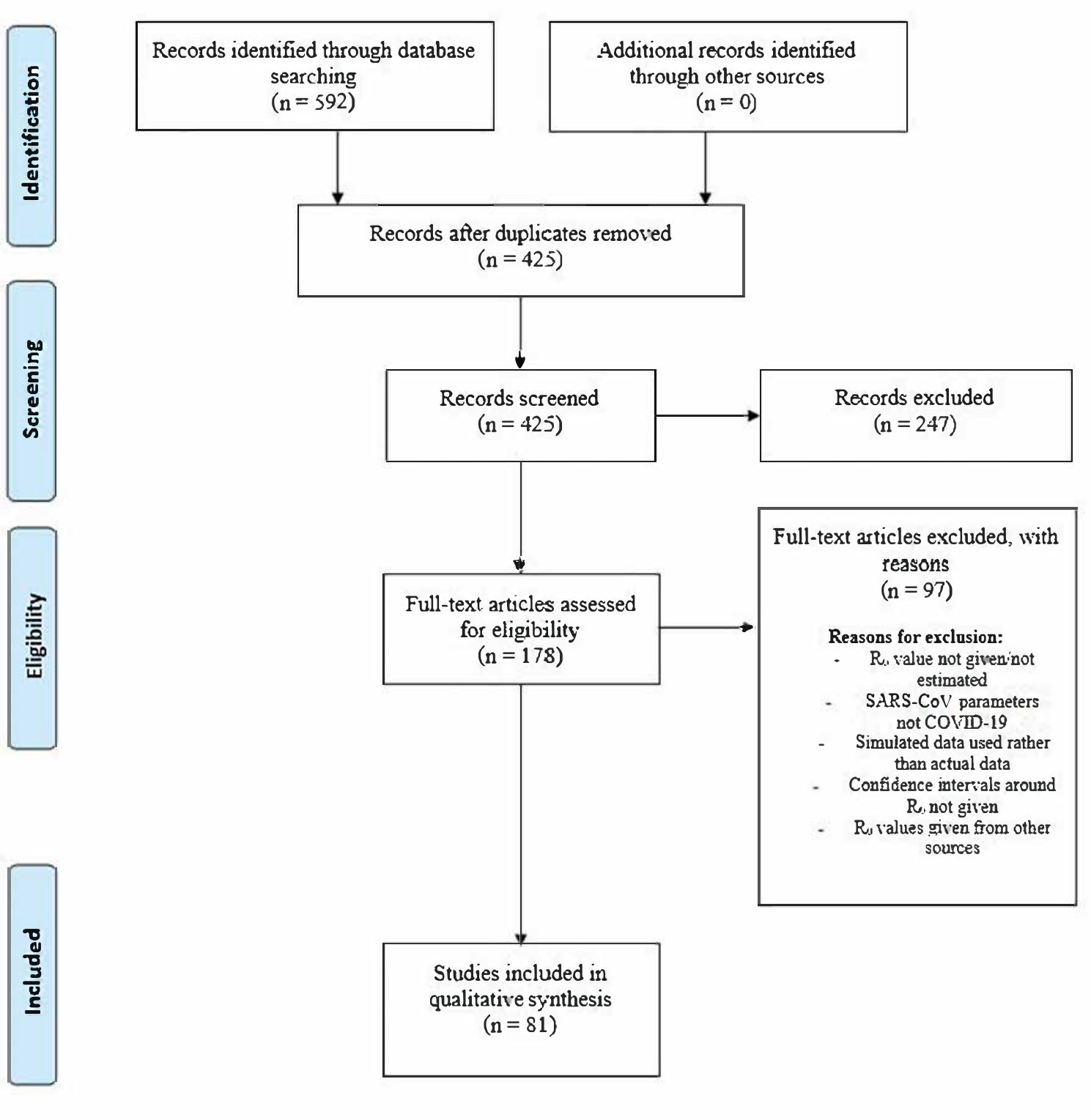}
    \caption{Article  screening and data extraction process for the systematic review \cite{moher2009} }
    \label{fig:PRISMA}
\end{figure}

\begin{figure}
	\begin{center}
		\includegraphics[scale=0.35]{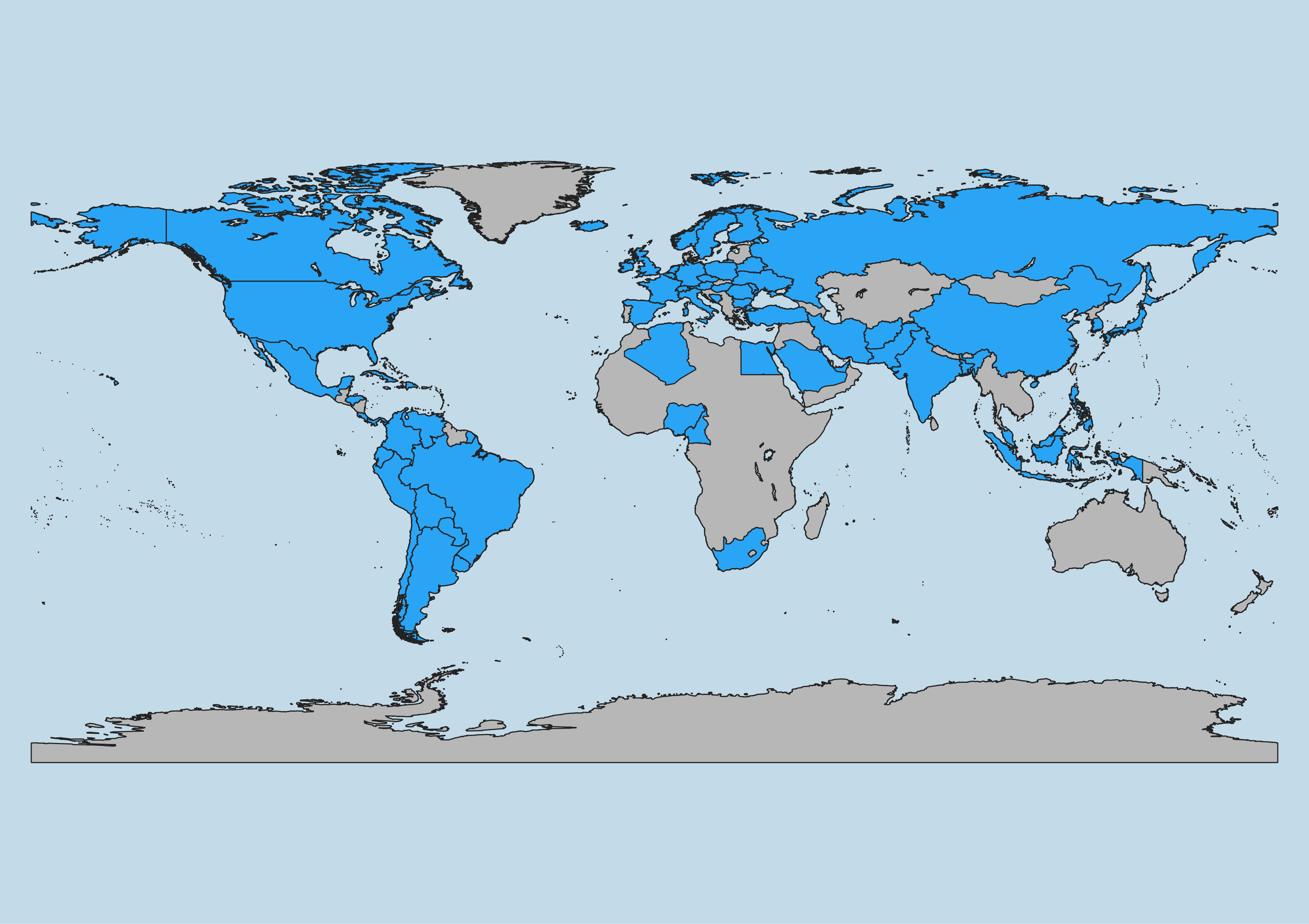}
	\end{center}
	\caption{Countries included in the study. \label{fig:WorldMap}}
\end{figure}

Figure \ref{fig:PRISMA} shows the review process. A total of 592 documents were found by searching PubMed, LitCOVID and the WHO COVID-19 database up to 30 June. In total, 425 were screened after removing duplicates, pre-prints and non-English articles. Of these, 81 were included in the study (Supplementary File 1). These contained sufficient methodology to calculate the basic reproduction number, as well as a numerical estimate for $R_0$. For each paper, the estimated values of $R_0$ were included, along with the country (countries) under study, the areas within countries where applicable, and the publication month. The studies covered 65 countries across 5 continents, as shown in Figure \ref{fig:WorldMap}.

Estimates for the basic reproduction number were given both at a national level and at local levels. The estimates for local areas were calculated independently and did not consider spatial auto-correlation between the regions. China had the most overall estimates at 100, followed by Italy with 91 and the USA with 61, thanks largely to the paper of \cite{peirlinck2020outbreak}, which provided $R_0$ estimates for each state of the USA. The paper of \cite{sarkar2020modeling} contributed to the large number of $R_0$ estimates for India, namely 27 estimates. Spain recorded the fifth largest number of estimates, namely 12. These estimates came from a greater variety of papers. The countries which recorded the highest number of national $R_0$ estimates were Italy with 17 estimates, China with 15, Japan with 11 and Spain with 10 national $R_0$ estimates. A total of 36 countries only recorded one national $R_0$ estimate each.

Since the focus herein is on spatial variation in $R_0$, we do not consider the estimation techniques. However, we note the most popular approaches observed in the study. SEIR-like compartmental models were used in two thirds (54) of the included papers, 15 papers employed exponential or logistic growth models, while the methods of \cite{cori2013new} and \cite{wallinga2004different} were used in 9 and 6 papers respectively. We refer to the review of \cite{liu2020reproductive} for a thorough investigation of the effects of estimation techniques on the $R_0$ estimate.

\subsection*{Variability in $R_0$ observed in the reviews}
\subsubsection*{Monthly distribution of $R_0$ value}

\begin{figure}
	\begin{center}
		\includegraphics[scale=0.6]{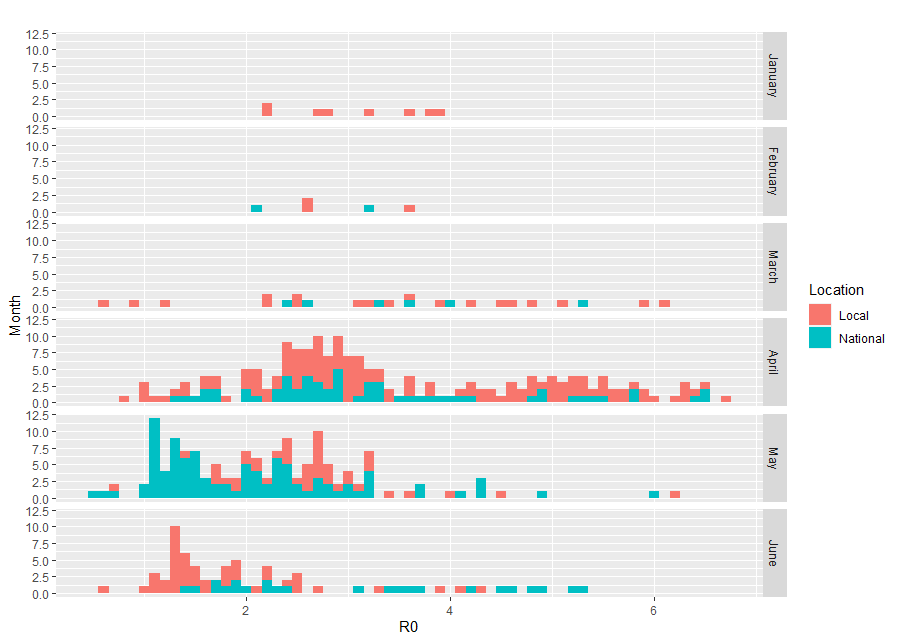}
	\end{center}
	\caption{Distribution of $R_0$ values by publication month. \label{fig:Histogram}}
\end{figure}

Figure \ref{fig:Histogram} shows the shift in $R_0$ estimates, both national and local, by publication month. For each month, the median was calculated based on all the papers that had been published up to and including that month. This does not indicate a change in the $R_0$ value over time, as most of the $R_0$ estimates were calculated based on data from the beginning of the pandemic. Rather, this reflects a change in the understanding of the disease over time. This may be due to various factors, such as more sophisticated estimation techniques or improving epidemiological knowledge. The estimates published in January and February were all obtained for China, and follow a similar pattern, being between 2 and 4. The estimates in January were all local, in particular calculated for Wuhan. For February, it should be noted that an estimate of size 14.8 for the Diamond Princess cruise ship is not visualised here, as it was removed from the analysis as an outlier. In March, estimates from the rest of the world became available, including Italy and Japan. A large amount of variability was experienced for this month. In April, 20 countries had estimates. Both local and national estimates evidenced a bi-modal distribution, with the greater mass concentrated between 1 and 3. The smaller concentration of mass, between 4 and 6, is largely due to the local estimates for the United States from the paper of \cite{peirlinck2020outbreak}. In May, the pattern shifted, with smaller $R_0$ estimates generally. Very few estimates were obtained above 4. By June, most estimates were between 1 and 2.5. The higher estimates were all obtained at a national level, with the top five recorded for Brazil, India, China, Spain, the UK and France. The histogram of all $R_0$ values is shown in Figure S1, exhibiting a bi-modal distribution similar to the patterns in Figure \ref{fig:Histogram}, with the greatest concentration of values between 1.5 and 3, and the second smaller mass between 4 and 6, mostly influenced by the local estimates published by \cite{peirlinck2020outbreak} in April.

\subsubsection*{National distribution of $R_0$ value}

\begin{figure}
	\begin{center}
		\includegraphics[scale=0.35]{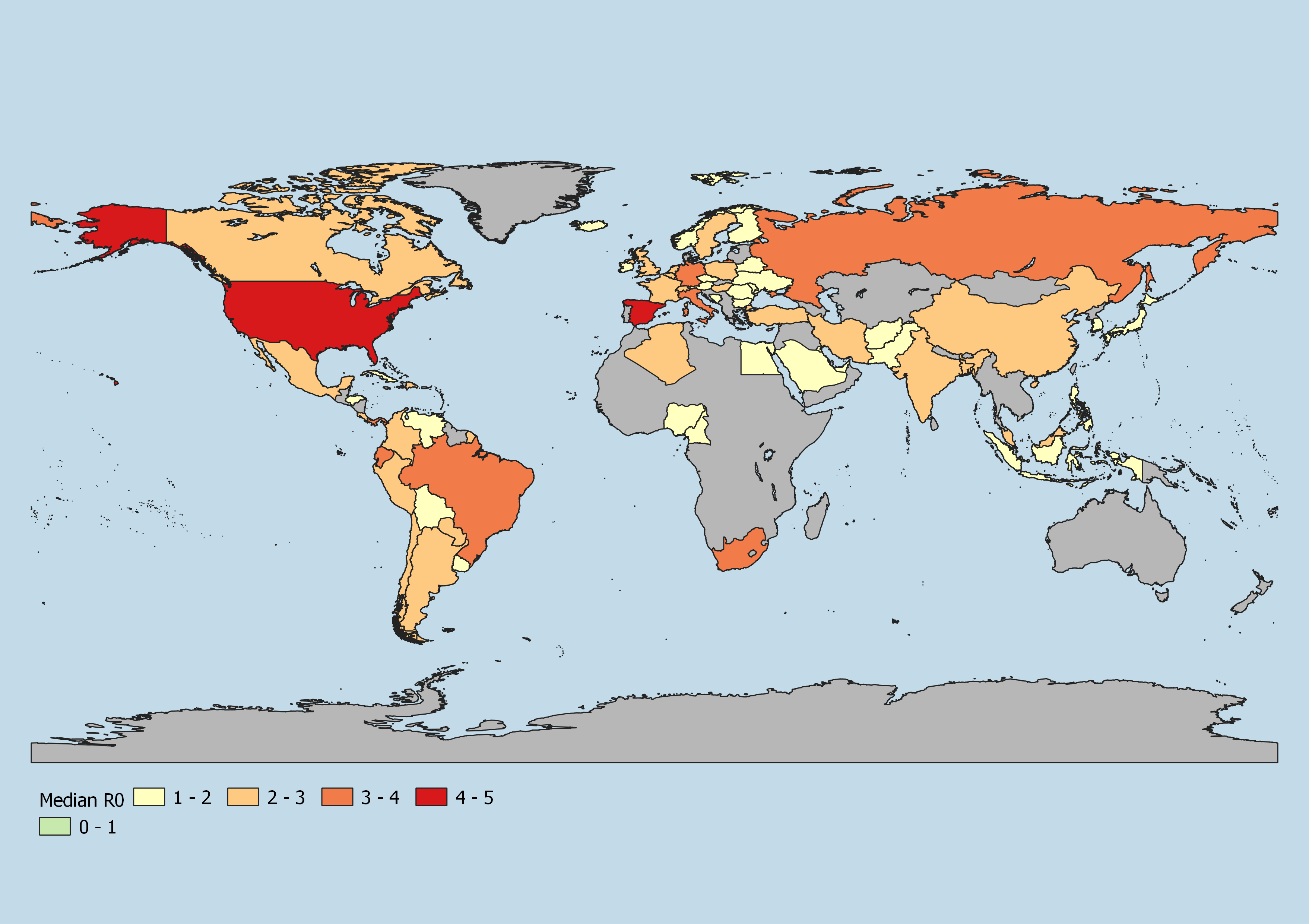}
	\end{center}
	\caption{National $R_0$ medians as of June 2020. \label{fig:NationalMedians}}
\end{figure}

Figure \ref{fig:NationalMedians} shows the median $R_0$ per country. As of June, most of the estimates were between 1 and 4, with only the USA (median $R_0$=4.78) and Spain (median $R_0$=4.25) being over 4. Only two were below 1, namely Hong Kong (median $R_0$=0.61) and Singapore (median $R_0$=0.7). Variance could not be measured at a national level due to the small number of estimates for many countries. Of the 65 countries, 36 had only one estimate available per country. The fact that many of these countries were included at all is  largely due to the contribution of \cite{lonergan2020estimates}, who provided $R_0$ estimates for 53 countries.

\subsubsection*{Regional distribution of $R_0$ value}

To investigate variability, the countries were aggregated, as variance can not be reliably calculated for the 36 countries with only one estimate. The countries were grouped into regions as defined by the World Bank\footnote{https://datahelpdesk.worldbank.org/knowledgebase/articles/906519-world-bank-country-and-lending-groups}. The groupings are shown in Table S1.

\begin{figure}
	\begin{center}
		\includegraphics[scale=0.35]{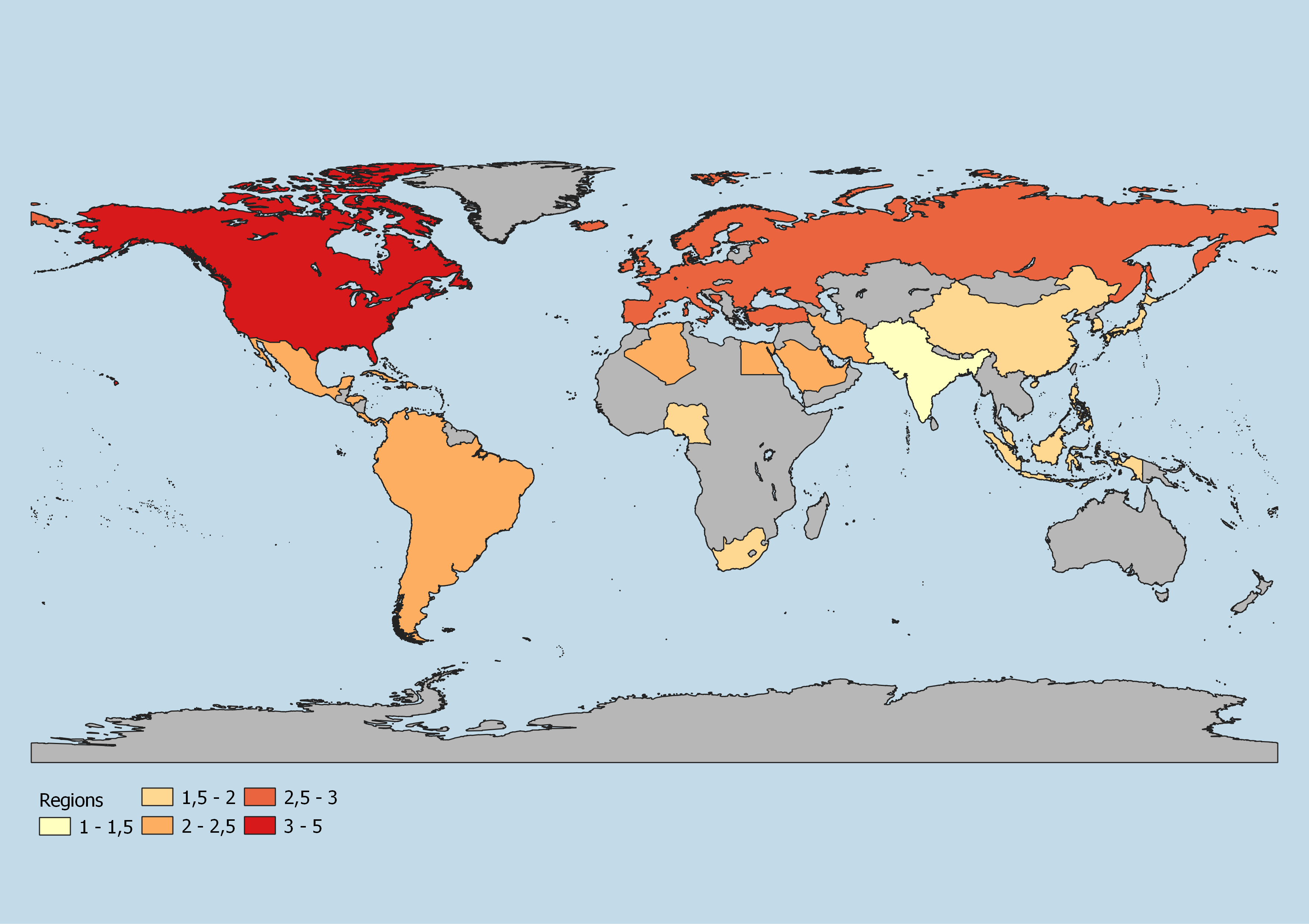}
	\end{center}
	\caption{Regional $R_0$ medians as of June 2020. \label{fig:RegionalMedians}}
\end{figure}

\begin{figure}
	\begin{center}
		\includegraphics[scale=0.35]{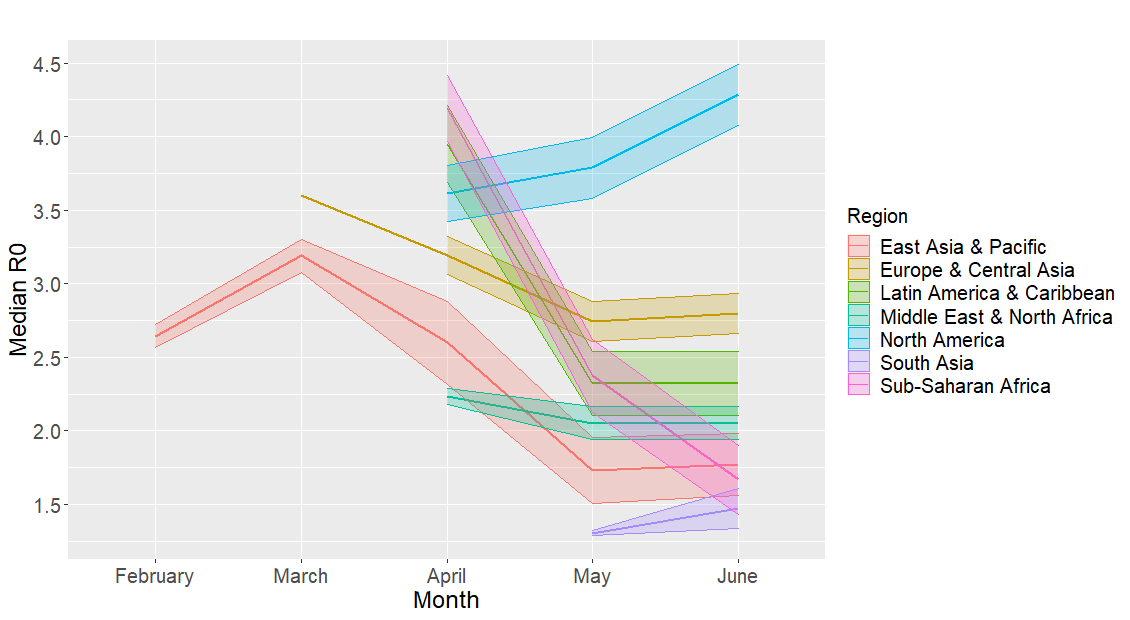}
	\end{center}
	\caption{Region medians over time. The bands show 10\% of the standard deviation. \label{fig:MedianPlot}}
\end{figure}

The regional median estimates are displayed on the map in Figure \ref{fig:RegionalMedians}, while Figure \ref{fig:MedianPlot} shows the change in regional medians by publication month of the estimates. The medians and standard deviations are cumulative, meaning they are based on all the estimates up to the relevant month. There is an initial increase for the East Asia \& Pacific region, from 2.65 in February (based on 2 estimates) to 3.19 in March. From March it decreases steadily to 1.77 in June. Similarly, the median decreases for all regions as shown in Figure \ref{fig:MedianPlot}, except for North America and South Asia. For these regions, the median $R_0$ increased. It is worth noting that neither of these regions had many estimates in earlier months. As of June, the region with the highest median estimate is North America at $R_0$=4.28, and the lowest median is obtained for South Asia at $R_0$=1.47. This places the $R_0$ estimate firmly above 1.

The bands in Figure \ref{fig:MedianPlot} show 10\% of the standard deviation, providing a visual representation of the variability per region. The 10\% is chosen for visual clarity, as showing the full standard deviation would cause the bands to overlap to the point of illegibility. The standard deviation increased in the East Asia \& Pacific region from February to March. This is due to the fact that only two estimates were available in February, namely 2.1 and 3.19, both from the same paper \cite{jung2020real}. By April, many more papers had been published, necessarily leading to a larger variance. By May, estimates began to stabilise, thereby reducing the variance. This general trend was observed for all regions for which more than two months were available, where an initial spike in the variance was observed for the second month, reducing in the third. For South Asia, only two months are available. It is therefore expected that the variance will decrease after June. As of June, the smallest variance was obtained for the Middle East \& North Africa region at 1.29 and the largest variance was obtained for Sub-Saharan Africa at 5.43. The variance also displays two distinct regional groups. The first experiences lower variance throughout, and includes Europe \& Central Asia, Middle East \& North Africa and South Asia. The second group exhibits higher variance, and consists of East Asia \& Pacific, Latin America \& Caribbean, and North America.

A large number of estimates was recorded for both Italy and China. China has a big influence on its region, as the median for East Asia \& Pacific drops from 1.77 to 1.52 when excluding China, and the variance drops from 4.46 to 1.59. Italy, however, does not have such a big influence on its region. Excluding Italy, the median and variance for Europe \& Central Asia are 2.67 and 2.01, respectively, as opposed to 2.8 and 1.85.

\subsubsection*{Social factors that affect the variability of $R_0$}

From the previous section, it is clear that the value of $R_0$ differs across countries and regions. Since the $R_0$ value was estimated at the beginning of the pandemic, this is not a result of varying laws or non-pharmaceutical interventions in the face of the pandemic, but rather some factors related to the ordinary state of affairs in a country prior to lockdown. A brief analysis is conducted to explore how $R_0$ varies based on social or mobility factors.

The United Nations Human Development Index\footnote{http://hdr.undp.org/en/content/human-development-index-hdi Accessed on 16 October 2020.} (HDI) is considered to measure social development, along with the Sum4All Sustainable Mobility Index\footnote{https://sum4all.org/ Accessed on 15 October 2020.} (SMI) as a transport index. The HDI measures the quality of life of a country’s population. It considers the length and health of citizens’ lives, their education and knowledge, and standard of living. It is easily interpretable and data is collected yearly, therefore the 2019 HDI was used herein. The SMI is a multi-faceted metric which measures how well a country is performing in terms of achieving sustainable transport for the future. It takes into account rural access, urban access, equality in the treatment of male and female users of the transport system, efficiency, safety, as well as air pollution and GHG emissions. The data for nearly all countries globally is publicly available on the Sum4All website. Countries were grouped into developing economies or transitioning/developed economies as classified by the UN \cite{un2020wesp}.

\begin{figure}[htp!]
	\begin{center}
		\begin{tabular}{cc}
			\includegraphics[scale=0.3]{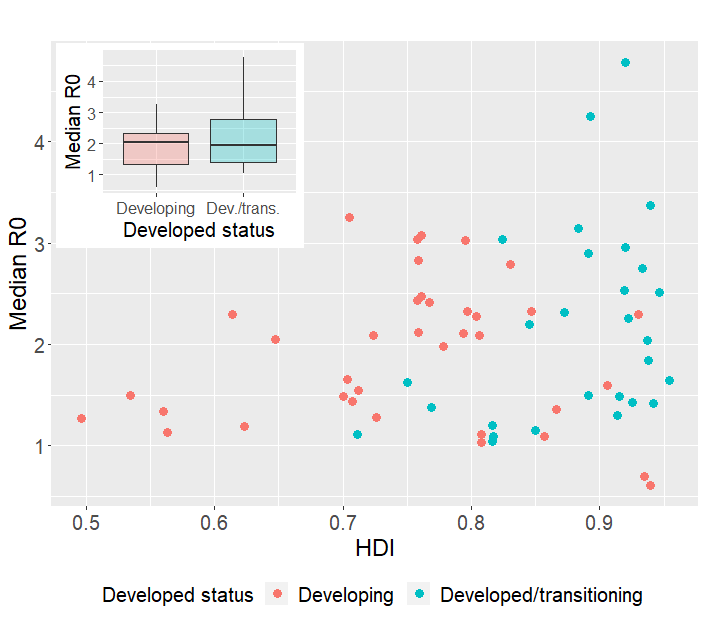} &
			\includegraphics[scale=0.3]{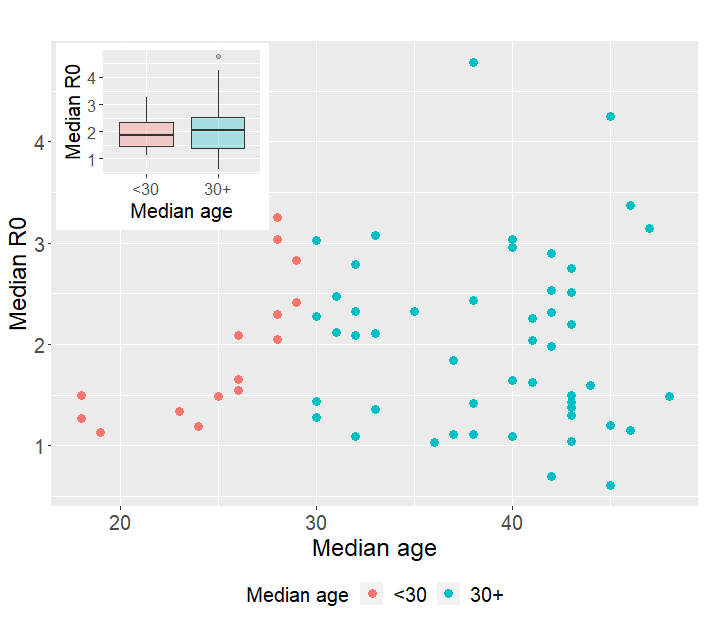} \\
			a) & b) \\
			\includegraphics[scale=0.3]{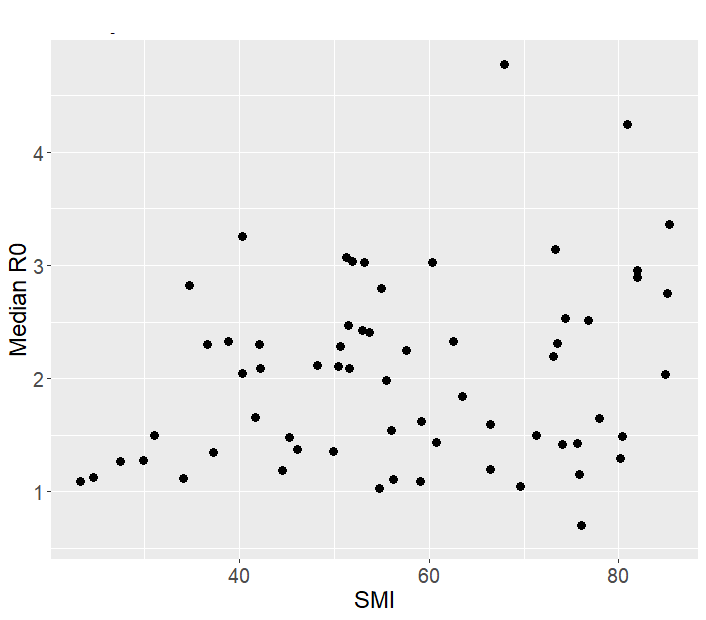} &
			\includegraphics[scale=0.3]{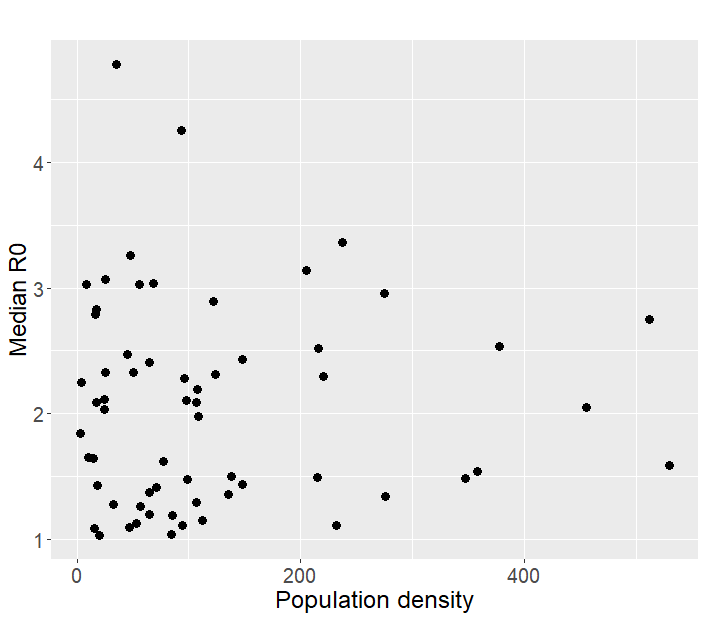}\\
			c) & d)
		\end{tabular}
		\caption{Relationship of $R_0$ to social factors. a) Median $R_0$ by HDI. Points are grouped according to developed-or-transitioning or developing economies. b) Median $R_0$ by median age. Points are grouped according to countries with a median age of 30 and over, or below 30. c) Median $R_0$ by SMI. d) Median $R_0$ by population density. For the purpose of the visualisation, countries with population density above 12 000 were excluded, namely Bangladesh, Singapore and Hong Kong.}
		\label{fig:Scatterplots}
	\end{center}
\end{figure}

Based on the $R_0$ values in the study, $R_0$ is not strongly correlated with HDI ($\rho=0.21$), SMI ($\rho=0.24$), median age ($\rho=0.08$), or population density ($\rho= -0.28$), indicating no linear relationships. Additionally, these variables are correlated with each other. This makes multivariate regression infeasible, however, trends in the data may be studied visually. Figure \ref{fig:Scatterplots} shows these patterns in the data. No high $R_0$ values were obtained for countries with low HDI, SMI or median age, or a high population density. Based on the available data, variation in $R_0$ increases as HDI, SMI and age increase, and decreases as population density increases, in perhaps a non-linear way. The boxplots show that the median $R_0$ is similar for developing versus developed countries, however the spread of values is negatively skewed for developing countries and positively skewed for developed countries. Additionally, the variability is higher for developed countries. When considering median age, a roughly symmetrical distribution was experienced for countries with a younger population, while the distribution of $R_0$ values was negatively skewed for countries with an older population. The latter category also experienced more variability than the former.

\subsection*{Illustrative example for South Africa}

South Africa's first case of COVID-19 was announced on 5 March 2020. The government took a swift risk averse approach placing the country in full lockdown due to high levels of co-morbidity in the population and concerns regarding hospital preparedness. We illustrate spatial variability of $R_0$ in South Africa as a special case of the variability observed in the systematic review to further guide the discussion of this paper.

\subsubsection*{Values of $R_0$ in South Africa}

There are three main approaches for estimating $R_0$ \cite{farrington2001estimation}, namely, using a mathematical expression of the parameters for contact amongst the population; using data of cases over time; or using expressions based on the values estimated from the endemic equilibrium of infection. An R package $R_0$ \cite{obadia2012r0} is also available to estimate the parameter for epidemic outbreaks, depending on data one has available. 

In South Africa, the National Institute For Communicable Diseases (NICD) has estimated $R_0$ (as well as the effective reproduction number) \cite{nicd2020}. They used the maximum likelihood method of \cite{forsberg2008likelihood} and the first 15 – 19 days of the epidemic in South Africa to estimate $R_0$. The main effect was due to international travellers entering South Africa at the start of the epidemic. Table \ref{tab:NICDR0} provides the estimates of \cite{nicd2020} for the most affected regions. The variability of the values indicates a non-homogeneous spreading of COVID-19 within South Africa. Accounting for spatial heterogeneity is therefore critical when modelling COVID-19 transmission, and $R_0$ in particular.

\begin{table}
	\begin{center}
		\caption{Estimates of $R_0$ in South Africa \cite{nicd2020}. \label{tab:NICDR0}}\vspace{0.4cm}
		\begin{tabular}{cccccc} 
			Region & South Africa & Western Cape & Gauteng & Eastern Cape & KwaZulu-Natal \\ 
            $R_0$ estimate & 2.07 & 1.76 & 2.19 & 1.84 & 1.73 \\ 
            95\% CI & (1.69, 2.5) & (1.11, 2.62) & (1.95, 2.39) & (1.1, 2.84) & (1.15, 2.47)
        	\tabularnewline 
		\end{tabular}
	\end{center}
\end{table}

\subsubsection*{Sensitivity analysis of spatial $R_0$}
\begin{figure}
    \centering
    \includegraphics[width = 9cm]{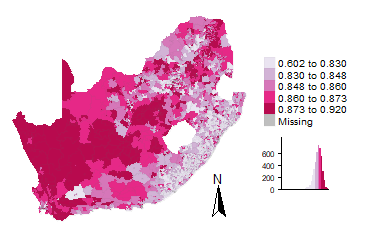}
    \caption{Graph of the correlations between $R_0$ and the proportion of infected individuals, at South African ward level as per the 2016 census.}
    \label{fig:sensitivitygraph}
\end{figure}

In order to demonstrate the importance of the spatial component, we conduct a local sensitivity analysis using a SEIR model for the COVID-19 cases in South Africa. We consider the sensitivity of the proportion of individuals that had been infected by the end of 201 days to the spatial\footnote{Here we conduct the spatial analysis at a ward level as per 2016 census data.} $R_0$. $R_0$ was sampled from a Uniform(1,5)\footnote{Based on the ranges observed in this systematic review.} distribution over 960 simulations\footnote{Chosen due to the number of cores available on the high performance computer made use of.}, and the correlation between the $R_0$ and proportion of infected individuals was calculated. Figure \ref{fig:sensitivitygraph} illustrates the results of this spatial sensitivity analysis visualised as correlations. 

The figure illustrates that many correlations are between 0.8 and 0.9, indicating a strong relationship between the spatial $R_0$ and the proportion of infected individuals. The figure further demonstrates that the strength of this relationship varies across geographical space.    This necessitates the use of a spatially varying $R_0$ to model COVID-19. Future work aims to provide such a spatial SEIR model for South Africa. 

\section{Discussion}
This study provided a comprehensive review of initial $R_0$ estimates of COVID-19, as modelled at the beginning of the pandemic in each country. The changes in the published estimates over the first half of 2020 displayed a growing understanding of the $R_0$ value in the modelling community. $R_0$ values were visualised at a national and regional level to investigate the spatial heterogeneity evident in the estimates. The relationship between $R_0$ and various social factors were visualised and described, and it was found that a higher $R_0$ value was related to higher values of HDI, SMI, and median age, and a lower population density. An illustrative example of COVID-19 modelling in South Africa highlighted the importance and the challenges of accounting for spatial heterogeneity when estimating $R_0$.

The study included 81 papers and covered 65 countries. The countries with the highest number of estimates, local as well as national, were China and Italy. The highest national $R_0$ medians were recorded in the USA (median $R_0$=4.78) and Spain (median $R_0$=4.25). The USA is well connected to other countries, including China, along direct flight routes and other channels of transport. This may have led to a higher influx of infected individuals before travel restrictions were put in place. A study by \cite{guirao2020covid} suggests that while the initial Spanish case was imported from Germany in late January, numerous cases were received from Italy through February before local infections began to rise by the end of February. Italy also recorded high estimates (median $R_0$=3.14). As it was one of the first countries to experience the pandemic, recording the second largest outbreak in March \cite{saglietto2020covid}, the country had little time to make preparations. This includes unofficial decisions by individuals to take preventative measures such as self-isolation and social distancing. While the values in this review were mostly estimated prior to official non-pharmaceutical interventions, citizens of countries that had more warning may have practised social distancing, self-isolation, and increased sanitising prior to official regulations. The value for Italy can therefore be seen as the most representative value of the basic reproduction number of COVID-19 in an uncontrolled environment. Aside from the USA and Spain, the only countries with a higher median $R_0$ than Italy were Germany (median $R_0$=3.37) and South Africa (median $R_0$=3.26).

The highest individual value of $R_0$ was 14.8 for the Diamond Princess cruise ship. This number may not be representative of COVID-19 transmission, as the disease spread rapidly in the confined conditions of the cruise ship. The influence of super-spreaders would also have been exacerbated by these conditions. Recent studies suggest that super-spreaders are responsible for the majority of infections \cite{majra2020sars,beldomenico2020superspreaders}.

The lowest median $R_0$ values were recorded for Hong Kong (median $R_0$=0.61) and Singapore (median $R_0$=0.7), the most densely-populated countries in the study. Both countries have a history of managing pandemics, so that unofficial non-pharmaceutical control measures, such as mask-wearing, may have been in place before official COVID-19 regulations were implemented. The lowest median $R_0$ values above 1 were recorded for Uruguay (median $R_0$=1.03), Romania (median $R_0$=1.04) and Saudi Arabia (median $R_0$=1.09). Note that a low $R_0$ does not necessarily mean fewer infections, since a low estimated $R_0$ might also be due to a lack of testing or unreliable case data. $R_0$ is usually estimated from the number of confirmed cases, which can be a direct result of a country’s testing strategy.

At a regional level, the highest regional median was recorded for North America (median $R_0$=4.28), while South Asia obtained the lowest median (median $R_0$=1.47). The median shows an increasing trend, with higher estimates for June than for May. This does not necessarily imply that the estimate will continue to rise. Most regions experienced a sharp decline in median $R_0$ from the first to the second month of estimation, with a slight increase into the third month. This indicates an initial overestimation by modellers, followed by an over-correction, finally stabilising to a slightly higher value that is nonetheless much lower than initial estimates. The exceptions to this trend were East Asia \& Pacific, which experienced a sharp peak before a decline, and North America, which increased steadily to a regional median of 4.28. It is worth noting that neither of these two regions had a large number of estimates. It is therefore possible that the median $R_0$ may decrease as more estimates become available, following the pattern of the East Asia \& Pacific Region. All other regional medians stabilised, showing a growing consensus among researchers that the $R_0$ value of COVID-19 is between 1 and 3.

Patterns in the variance differed by region. East Asia \& Pacific had a low initial variance, calculated on two estimates for China in February. This increased slightly into March, spiked to a value of nearly 8 in April, then decreased sharply in May and more gradually in June. For the Middle East \& North Africa, Europe \& Central Asia, and North America, the variance in the published estimates showed a spike in uncertainty from April to May followed by stabilisation in June. Sub-Saharan Africa followed a similar trend, experiencing a greater decrease from May to June. The above mentioned regions all follow a similar pattern to East Asia \& Pacific, lagged by two months. This may be due to later outbreak dates. Latin America \& Caribbean shows a sharp decrease from the initially high variance (close to 7) in April, to a variance below 5 in May, and a slower decrease in June. This follows precisely the trend in East Asia \& Pacific, not lagged. The variance for South Asia increased sharply from May to June. Based on the trends of the other regions, we may expect this variance to decrease in July.

The impact of various social and demographic factors on $R_0$ was studied. In particular, the HDI and development level were used as indicators of socio-economic development, while the SMI indicated the level of development of the transport network of each country. The median age and population density were considered as demographic factors. While no linear relationships were found between $R_0$ and any of these factors, clear patterns were visible. In general, there were no high median $R_0$ values for countries with low measures of social development and infrastructure, i.e. a developing economy, low HDI or low SMI. It should be noted that most estimates were obtained using the data of reported cases. In the case of developing countries, a substantial portion of the population may not have access to testing facilities due to lack of infrastructure, leading to fewer cases being detected. No countries with younger populations had high median $R_0$ values. In contrast, no low median $R_0$ values were recorded for countries with a high population density. Additionally, the data showed that the variability in $R_0$ depended on the value of these factors. Variability and hence uncertainty in $R_0$ increases with age, HDI and SMI, and decreases with population density. 

The South African example illustrates the necessity of accounting for spatial heterogeneity when modelling $R_0$, while highlighting the difficulty of modelling $R_0$ in real-time. The $R_0$ values estimated by the NICD illustrate that the spread of COVID-19 varies by geographical region within South Africa. This supports the findings of the literature review that $R_0$ varies spatially. The sensitivity analysis further demonstrated a spatially varying relationship between $R_0$ and the proportion of infected individuals, with differing correlations observed per ward. Accounting for spatial variation in $R_0$ is crucial for modelling the spread of COVID-19. This must however be done with caution, as adding a spatial component to a model will increase complexity and may increase uncertainty. A thorough understanding of the ways in which $R_0$ varies spatially is therefore required. This study provides a critical stepping stone towards such an understanding.

As more data becomes available, future work should incorporate $R_0$ literature of later months, including the rest of 2020 and beyond. In addition, this would allow further modelling to better quantify the relationship of $R_0$ to the social and demographic factors. The date of the initial outbreak per country could also be included, as well as unofficial social prevention measures put in place by citizens prior to official lockdown, as these may affect the value of $R_0$. Estimation techniques could also be considered, as in the review of \cite{liu2020reproductive}. Since $R_0$ is heavily reliant on the testing strategy of a country, testing statistics per country may also be included, subject to data availability. Lastly, the methodology of this study could be extended to investigate spatial variation of the time-varying reproduction number.

\section{Conclusion}
This systematic review presented a summary of the $R_0$ estimates of COVID-19 across the first six months of the pandemic, indicating a changing understanding of the disease characteristics. Spatial variation in the estimates was presented at a national and regional level. In addition, the study explored the relationships between $R_0$ and development and SMI. Countries with a lower level of development, a younger population, a lower sustainable mobility score or a higher population density did not record high $R_0$ values. The example for South Africa demonstrated the necessity of introducing a spatial component into transmission modelling. This study provides a vital step towards characterising spatial variation in $R_0$. A deeper understanding of the factors that influence the spread of COVID-19 is critical for combating this disease which has disrupted so many lives across the globe.

\section*{Conflict of interest}
The authors have no conflict of interest to declare for this study.

\section*{Acknowledgements}
This work is supported by the NRF-SASA (National Research Foundation-South African Statistical Association) Crisis in Statistics Grant. We acknowledge the use of the support of the Center of High Performance Computing.  The contributions of and discussions with Sally Archibald, University of the Witwatersrand, are also acknowledged. 

\section*{Author contributions}
Conceptualization: IFR, RT, NA, RMD, JH, CJVR, PD, NDT, ZK, ALR. 
Data curation: RT, NA, IFR.
Formal analysis: RT, NA. 
Funding acquisition: -
Methodology: NA, RT, RMD, JH.
Project administration: IFR, PD. 
Visualization: RT, RMD.
Writing – original draft: RT, NA, IFR, RMD, ZK. 
Writing – review \& editing: IFR, RT, NA, RMD, JH, CJVR, PD, NDT, ZK, ALR. 

\bibliographystyle{vancouver}
\bibliography{R0Paper1Bibliography.bib}

\end{document}